Core Components of Emotional Impulsivity: A Mouse-Cursor Tracking Study

Anton Leontyev[1], Takashi Yamauchi[2]

[1]Geneva College, [2]Texas A&M University

**Author Note**

Anton Leontyev https://orcid.org/0000-0001-8880-6161;

Takashi Yamauchi http://orcid.org/0000-0002-6372-1118.

The data and syntax for this project are publicly available at https://osf.io/wgu6p/.

Word count: 11264

The authors report no conflicts of interest.

Please address correspondence regarding this article to Anton Leontyev at axleonty@geneva.edu

**Abstract**

Impulsive individuals exhibit abnormal reward processing (heightened preference for immediate rewards, i.e., impulsive choice, IC) and a penchant for maladaptive action (the inability to inhibit inappropriate actions, i.e., impulsive action, IA). Both impulsive choice and impulsive action are strongly influenced by emotions (emotional impulsivity); yet how emotions impact impulse behavior remains unclear. The traditional theory suggests that emotions primarily exacerbate impulsive action and prompts impulsive choice. The alternative theory states that emotions primarily disrupt attention (attentional impulsivity, AImp) and prompt impulsive choice. However, the empirical evidence supporting these theories is inconsistent—few correlations have been reported between self-report measures of emotional impulsivity and behavioral measures of impulsivity beyond clinical populations. In two studies, we probed the interplay among emotions, impulsive action (IA), attentional impulsivity (AImp), and impulsive choice (IC). We elicited positive and negative emotions using emotional pictures and examined the extent to which elicited emotions altered behavioral indices of impulsivity. Our findings suggest that, in a nonclinical population, positive and negative emotions exacerbate both impulsive action and attentional impulsivity in a different manner. While negative emotion increases impulsive action, both positive and negative emotions heighten attentional impulsivity, suggesting that positive and negative emotions play diverging roles in impulsive behavior.

Core Components of Emotional Impulsivity: A Mouse-Cursor Tracking Study

Cognitive control—the ability to override one's impulses and make decisions based on one's goals rather than habits or reactions—is perhaps one of the most distinctive characteristics of human cognition (Stout, 2010). Deficits in cognitive control—impulsivity—affect the quality of life universally (Victor et al., 2011). Impulsivity is a central component of mental disorders, including attention-deficit/hyperactivity disorder (ADHD; Winstanley et al., 2006), borderline personality disorder (BPD; Sebastian et al., 2013), gambling addiction (Verdejo-García et al., 2008), as well as drug abuse (Perry and Carroll, 2008), smoking, and alcoholism (Granö et al., 2004).

Impulsivity includes at least three different facets: (1) impulsive choice (IC), also known as "delay discounting" and overwhelming preference for smaller immediate rewards over larger delayed ones (MacKillop et al., 2016); (2) impulsive action (IA), also known as response inhibition (Horn et al., 2003) and troubles with suppressing inappropriate behavior (MacKillop et al., 2016); (3) attentional impulsivity (AImp), inability to focus and maintain attention (Cservenka and Ray, 2017).

IC and IA are considered two primary facets of the general construct of impulsivity (Sharma et al., 2014); elevated levels of IC and IA are found in many disorders, including ADHD and addictive disorders. When impulsivity is studied through self- or other-reports, impulsive individuals often show heightened impulsive choices and impulsive actions. However, when impulsivity is studied using behavioral/cognitive tasks (e.g., delay discounting tasks and stop-signal tasks), a wide discrepancy emerges. For example, individuals with strong impulsive choice tendencies show little impulsive actions (Broos et al., 2012). Likewise, attentional impulsivity is not correlated with either impulsive action (Khng and Lee, 2014) or impulsive choice (Martinez-Loredo et al., 2017) in populations without clinical impairments.

A component of impulsivity that is frequently observed in impulsive choice, impulsive

action, and attentional impulsivity is emotions (Figure 1a). As Johnson and colleagues (2017)

show, the tendency to respond impulsively in the presence of emotions (emotional impulsivity)

constitutes a distinct form of impulsivity. Recent studies suggest that emotions are crucial for understanding the psychopathology of impulsive disorders (Johnson et al., 2013), as well as daily impulsivity-related behavioral problems, e.g., aggression and substance abuse (Sharma et al.,

2014). However, the specific neurocognitive mechanism by which emotions influence impulsive behaviors is unclear.

**Figure 1**

*Emotions and Impulsive Behaviors*

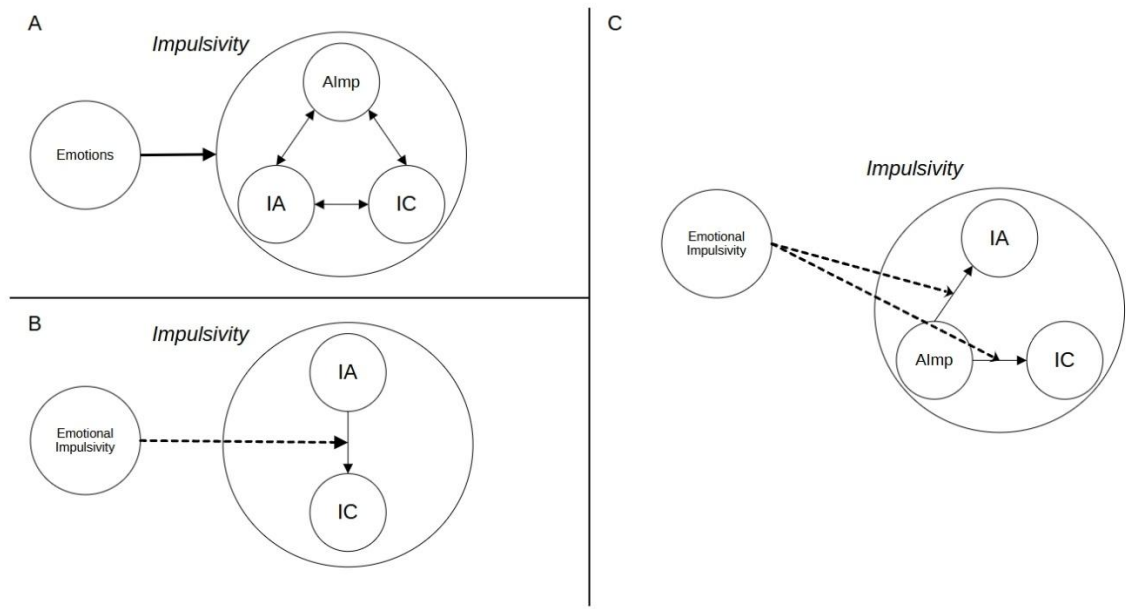

The traditional theory posits that impulsive action (e.g., troubles with suppressing inappropriate behavior) is the main cognitive trait that underlies emotional impulsivity (Johnson et al., 2020). As case in point, the activity in the serotonergic system – the primary neural circuit that underpins impulsive behaviors – is shown to be linked to response inhibition (Homberg, 2012). However, this view is challenged by emerging evidence. First, recent studies do not replicate the predicted correlation between response inhibition performance and emotional impulsivity scores (Elliott et al., 2022). Second, the evidence for response inhibition is mainly found in the clinical population, where it might stem from general cognitive deficits (Aichert et al., 2012). Third, serotonergic malfunction does not exclusively influence response inhibition; it also impacts other forms of executive functions such as attention (Wingen et al., 2008).

Alternative theory suggests that attentional impulsivity is a main cognitive trait underlying emotion-related impulsivity (Rosenthal et al., 2024). First, attentional impulsivity is often comorbid with emotional dysregulation (Jakubczyk et al., 2018). Second, emotional stimuli restrict attention to themselves; impulsive choices and impulsive actions commonly accompany attentional bias (Pessoa, 2008). However, the direct correlations between attentional impulsivity and emotional impulsivity are weak (Malivore et al., 2019).

This study investigates the specific mechanism by which emotions influence impulsive behaviors; in particular, we examine whether impulsive action/poor response inhibition (Figure 1b, upper panel) or attentional impulsivity (Figure 1c) is the main cognitive component (cognitive trait) of emotional impulsivity.

**Emotions influence impulsive action (IA) and attentional impulsivity (AImp)**

Sharma and colleagues' (2014) meta-analysis shows that most self-report impulsivity measures loaded onto two affect-related factors: conscientiousness-like and extraversion-like, implying that most impulsive behaviors are driven by affect. Some researchers hypothesize that emotion exacerbates response inhibition (impulsive action) (Chester et al., 2016), while others hypothesize that emotion interferes with attention (Pessoa, 2008).

The connection between emotional impulsivity and impulsive action is evident on behavioral and neural levels. For example, patients with schizophrenia and a history of violence tend to show less activation in inhibition-related regions of the brain (e.g., dlPFC) and make more commission errors during an emotional go/No-go task compared to non-violent individuals with schizophrenia (Tikasz et al., 2018). Although the correlations between EI and IA are stronger in clinical samples (mean $r = .34$), they are much weaker (mean $r = .14$) in community samples (Johnson et al., 2020), suggesting that the relationship between emotional impulsivity and response inhibition might result from a general cognitive deficit and not a specific IA-EI relationship, as Johnson and colleagues suggest (Johnson et al., 2017).

There is also inconclusive evidence about the strength of the relationship between emotional impulsivity and impulsive action when an affect (negative or positive) is not present. Jauregi (2018) showed no significant relationship between measures like SSRT (stop-signal reaction time, a common index of the inhibitory ability, Verbruggen and Logan, 2008) and negative urgency, as measured by UPPS (Urgency-Premeditation-Perseverance-Sensation Seeking-Positive Urgency scale, a questionnaire for emotional impulsivity) but found significant differences in UPPS scores between low- and high-impulsivity groups. In contrast, strong associations between emotional impulsivity and performance in the go/No-go task or the stop-signal task even when mood induction fails (Gunn and Finn, 2015) or report no interaction with mood strength (Dekker and Johnson, 2018). Yet other studies report a strong relationship between emotional impulsivity and behavioral inhibition measures in stop-signal and other inhibition tasks (e.g., Flanker task; Gabel and McAuley, 2018, 2022) when an affect is present.

A similar ambiguity is evident in the relationship between emotional impulsivity and attentional impulsivity. For example, emotional stimuli in a Stroop task were found to exacerbate attention (Strauss et al., 2005); Dysregulation of emotions frequently accompanies inattention in impulsivity-related disorders such as ADHD (O'Neil & Rudenstine, 2019; Rosen et al., 2015). Positive emotional impulsivity (i.e., tendency to react impulsively under positive emotions) was correlated with Stroop task performance (Sharma et al., 2014). However, many other tasks and questionnaires show little to no correlation between emotional impulsivity and inattention measures, particularly negative urgency (Becker et al., 2016). It should also be noted that the correlation between impulsive choice and emotional impulsivity rarely replicates in non-clinical populations (Burnette et al., 2019; Jauregi et al., 2018), although Jauregi and colleagues (2018) found significant differences between high and low emotional impulsivity groups by comparing their proportions of larger/later vs. sooner/smaller choices in the delay discounting task (DDT).

Similarly, neuropsychological evidence remains inconclusive. The individual differences in impulsivity are attributed to the amount of serotonin available in the brain (serotonergic function; Carver and Johnson, 2008). Higher serotonergic function facilitates better response inhibition when these impulses are triggered by the emotional stimuli (impulsive action, as predicted by Johnson and colleagues view); the amount of serotonin primarily affects an individual's penchant for impulsive action, as evidenced by lesion studies (Logue and Gould, 2014). Similarly, depleting serotonin led to more immediate choices in a delay discounting task (impulsive choice), whether emotional stimuli were present (Bari et al., 2010; see Puig Pérez, 2018 for review) or not (Schweighofer et al., 2008; Crockett et al., 2010). However, serotonin is also strongly related to the allocation of attention. Primate studies show that depleting serotonin also leads to reduced stimulus looking times, i.e., inattention (Weinberg-Wolf et al., 2018).

Individuals with a chronic deficit of serotonin (5-HT) show worse performance in an attentional task (Banerjee and Nandagopal, 2015). Variation in the serotonin transporter gene is associated with attentional biases towards positive and negative emotional stimuli (Fox et al., 2009). Depleting serotonin also leads to lapses in attention (Zepf et al., 2010).

In sum, emotional impulsivity is closely related to impulsive action (IA) and attentional impulsivity (AImp). However, the primary cognitive trait underlying emotional impulsivity remains unclear.

**Methodological Challenges and Present Studies**

Because impulsivity likely manifests on multiple fronts (Hampton et al., 2015), it is difficult for a single performance-based metric like stop-signal reaction time (SSRT; Verbruggen and Logan, 2008) to capture its varying facets (Caswell et al., 2015). Furthermore, since behavioral tests such as stop-signal tasks and go/No-go tasks have low test-retest reliability (Enkavi et al., 2018; Hedge et al., 2019), it is difficult to study emotional impulsivity with nonclinical populations. Indeed, much research demonstrates that performance-based behavioral tests and questionnaire-based self-reports rarely correlate due to their structural discrepancies (Dang et al., 2020; Wennerhold et al., 2020; Toplak et al., 2013).

To redress these limitations, in this study we applied a recently developed action-based measure of behavior—"mouse-cursor tracking." In a mouse-cursor tracking task, participants maneuver the computer cursor to select a response button; by analyzing the cursor's navigational path, trajectory features such as velocity and acceleration reveal participants' perceptual, cognitive, and social conflicts in the decision-making process.

This approach has been successfully employed to investigate decision-making processes across various contexts, including perceptual judgment, semantic categorization, linguistic judgment, and social judgment (Chapman et al., 2010; Dale et al., 2007; Farmer et al., 2007; Freeman & Ambady, 2009; Freeman et al., 2010; Freeman, 2018; Song & Nakayama, 2008; Schneider et al., 2015; Spivey et al., 2005; Stillman & Ferguson, 2019; Xiao & Yamauchi, 2014, 2015, 2017; Xiao et al. 2023; Yamauchi et al., 2007). Mouse-cursor tracking has demonstrated relevance to emotional states and attitudes, such as ambivalence, anxiety, and mindsets (Yamauchi & Xiao, 2018; Yamauchi, 2018; Yamauchi et al., 2019). Recent studies have shown that features of mouse movement (and hand motion), such as peak velocity and acceleration, are valid indicators of impulsivity (Leontyev & Yamauchi, 2019, Leontyev et al., 2018, Leontyev et al., 2019, Leontyev & Yamauchi, 2021; Yamauchi et al., 2024; Shadmehr & Ahmed, 2020).

The purpose of present studies is to *experimentally* test the "Core Impulsive Action" and "Core Attentional Impulsivity" theories of emotional impulsivity with a nonclinical population. Because there are few theoretical frameworks to link actual motor measures to impulsivity, Study 1 first identifies behavioral indicators, patterns of mouse-cursor motion, that are associated with impulsive action (IA), attentional impulsivity (AImp), and impulsive choice (IC), respectively. In Study 2, we elicited positive, negative, and neutral emotions in individual participants using standardized emotional pictures and investigate the extent to which these elicited emotions alter participants' mouse-cursor motions.

If the core deficit in emotional impulsivity is impulsive action ("Core Impulsive Action theory"), behavioral indicators of impulsive action should deteriorate in trials that embed emotional stimuli compared to those with emotionally neutral stimuli. Conversely, if the core deficit in emotional impulsivity is inattention ("Core Attentional Impulsivity theory"), behavioral indicators of attentional impulsivity should deteriorate in trials that embed emotional stimuli compared to those with emotionally neutral stimuli.

**Study 1: Identifying behavioral indicators of impulsive action, attentional impulsivity, and impulsive choice**

To evaluate the two divergent accounts of emotional impulsivity, it is essential to identify mouse-cursor movement features, such as peak velocity, acceleration, and entropy (Figure 3), that are associated with three latent constructs of impulsivity: impulsive action (IA), impulsive choice (IC), and attentional impulsivity (AImp). To do so, we conducted a confirmatory factor analysis (CFA; Jacobucci et al., 2019) to identify mouse-cursor movement features that cluster with impulsivity metrics derived from self-report questionnaire. We then used these metrics to compare "Core Attentional Impulsivity" and "Core Impulsive Action" theories on a trait level.

For impulsive action, we initially selected BIS-M (Barratt Impulsiveness Scale– Motor subscale) as the major index of impulsive action and included additional behavioral measures (commission error and stopping distance in the stop-signal task) that clustered with BIS-M. For attentional impulsivity, we initially selected BIS-Attentional Impulsivity (BIS-A) scores and CAARS-Inattention (Conners' Adult ADHD Rating Scales, subscale A) scores as the major indices of impulsive attention and included additional behavioral measures (SD of maximum velocity in "go" trials) that clustered with BIS-A and CAARS-A. Our choice of behavioral measures for attentional impulsivity was based on Unsworth and Miller's findings, indicating that the distance and time of mouse movement are indicative of individual differences in attentional control (Unsworth & Miller, 2024). For impulsive choice, we selected delay discounting rate $k$ and choice consistency parameter $\beta$ obtained in the delay discounting task (Wilson & Collins, 2019; Odum, 2011), as well as BIS-Nonplanning (BIS-N) scores. For emotional impulsivity, we selected UPPS-Negative Urgency and UPPS-Positive Urgency (Whiteside & Lynam, 2001). All calculations were performed using R package *lavaan* (Rossell, 2012).

In identifying behavioral indicators (mouse-cursor movement measures), we focus on the velocity and the spatial deviation of mouse-cursor motions as these motion features are known to provide insights into the mental states of the decision maker (Shadmehr & Ahmed, 2020; Stillman et al., 2018; Stillman & Ferguson, 2019; Leontyev & Yamauchi, 2021). Specific mouse-cursor movement measures we included for our analysis are explained later in the Method section.

Assuming that these pre-selected measures of impulsivity represent latent constructs of impulsive action, impulsive choice, and attentional impulsivity, respectively, our goal here is to identify behavioral indicators of impulsivity—i.e., mouse-cursor movement measures obtained from the stop-signal task and the delay discounting task— that align with impulsive action, impulsive choice, and attentional impulsivity.

**Methods**

***Transparency and Openness***

In the sections that follow, we report all manipulations and all measures in the study. All data are available at https://osf.io/wgu6p/?view_only=72958acaba264690bbf2a2100349672d/. Data were analyzed using R, version 4.2.3 (R Core Team, 2022). Study 1's design and its analysis were not pre-registered. This study has been approved by the Institutional Review Board of Texas A&M University (approval ID: IRB2021-0908D).

***Participants***

A total of 298 individuals were recruited for this study from the Amazon M-Turk website (location limited to the United States of America). Only individuals who managed to attain at least 5% accuracy in the primary task (explained later in the next section) and successfully inhibit their responses in at least 5% of "stop" trials (commission error) were included in the final sample. We chose this threshold because including more trials improves the reliability of inhibitory indices (e.g., SSRT; Congdon et al., 2012), particularly when healthy adult samples are considered[1]. Moreover, only participants who indicated their control device as a mouse were selected. The final sample included 200 individuals (mean age 44.03, SD = 12.33). Of these participants, 111 indicated their gender as female (mean age 46.60, SD = 12.10), 88 as male (mean age 40.90, SD = 11.90), and one as "other" (26 years).

*Procedure*

After completing the virtual consent form, participants carried out the stop-signal task and the delay discounting task, and answered impulsivity questionnaires (BIS, UPPS-P, and CAARS). All studies were programmed in the *PsychoPy* environment (Peirce et al., 2019).

**Stop-signal task (SST).** The stop-signal task in this study followed the procedure outlined by Ma and Yu (2016). Participants were exposed to a series of 200 trials, wherein they were presented with a random dot kinematogram (Scase et al., 1996) of 100 dots, employing a limited lifetime algorithm. Among these dots, a certain proportion (10, 50, or 80%) moved coherently either left or right while the remaining dots moved in random directions. The specific ratio of coherent dots was determined randomly at the beginning of each trial.

---

[1] We have also applied more stringent criteria (i.e., at least 20% and 40% accuracy in primary and stop-signal tasks). The results of these analyses remained largely the same.

**Figure 2**

*Stop-signal Task*

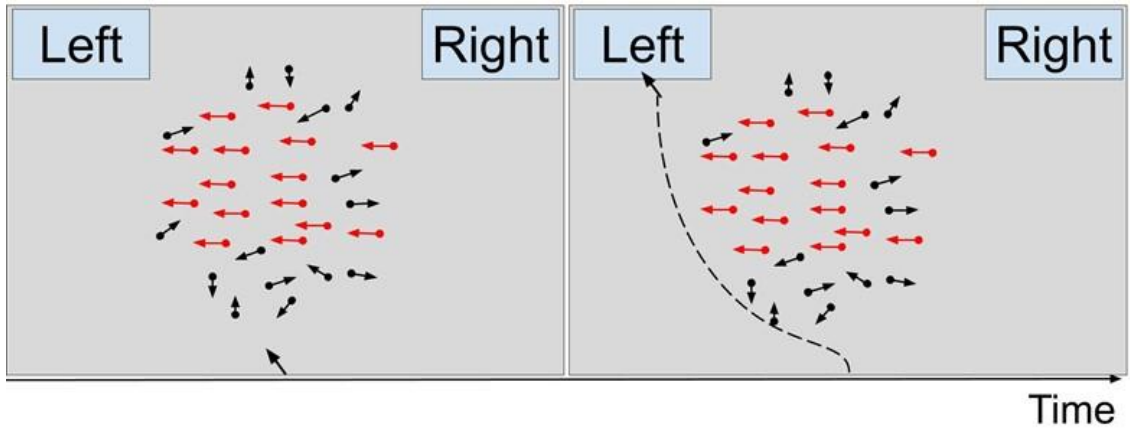

*Note*: In the mouse movement version, the response is made by moving a mouse and clicking on a response box. Red arrows indicate coherent dots.

The primary task, referred to as the direction discrimination task (Figure 1), was for participants to indicate the direction in which the coherent dots were moving. This was accomplished by clicking on a button drawn on the screen during "go" trials. In 25% of the trials (50 out of 200), participants were presented with an auditory stop-signal ("stop" trials). Before the study, participants were instructed to cease their response if they hear the stop-signal. The delays after which the stop-signal is delivered were chosen randomly and uniformly from a predetermined set: 100, 200, 300, 400, 500, or 600 ms.

To align the starting position between trials, participants were required to click a start button at the beginning of each trial. The mouse cursor was placed at the bottom-center of the screen (0, -0.8) at the beginning of each trial, where (0,0) is the center of the screen and (1, 1) denotes the top right corner of the screen. The mouse cursor coordinates were recorded every 16 ms, starting at the beginning of the trial and continuing until either 3000 ms elapsed or a response was made (Leontyev and Yamauchi, 2019). The measures collected in the stop-signal task included commission error as well as mean and standard deviation of response time in "go" and "stop" trials. Participant had up to 3000 ms to initiate movement; the response was made by clicking on one of the buttons drawn on screen.

Based on commission errors and response times, we calculated the stop-signal reaction time (SSRT) – a common index of the inhibitory ability – using the integration method (Verbruggen and Logan, 2008). Longer SSRTs are associated with ADHD (Crosbie et al., 2013), borderline personality disorder (Sebastian et al., 2013), and Parkinson's disease (Gauggel et al., 2004). The integration method estimates the SSRT by "integrating" the RT distribution and finding the point at which the integral is equal to the probability of response given that a stop-signal is present.

**Delay discounting task (DDT).** The present study employed the delay discounting task (DDT) adapted from MacKillop et al. (2016) (Figure 3). In this task, participants were required to make a hypothetical monetary choice between an immediately available smaller reward and a larger reward available after a delay. Delay for the smaller/sooner option was held constant (immediately), while rewards were randomly and uniformly selected from $10, $20, $30, $40, $50, $60, $70, $80, $90, or $99. For the larger/later option, the reward amount was held constant at $100, while delays were chosen from 1, 7, 14, 30, 60, 90, 180, or 365 days.

**Figure 3**

*Delay Discounting Task*

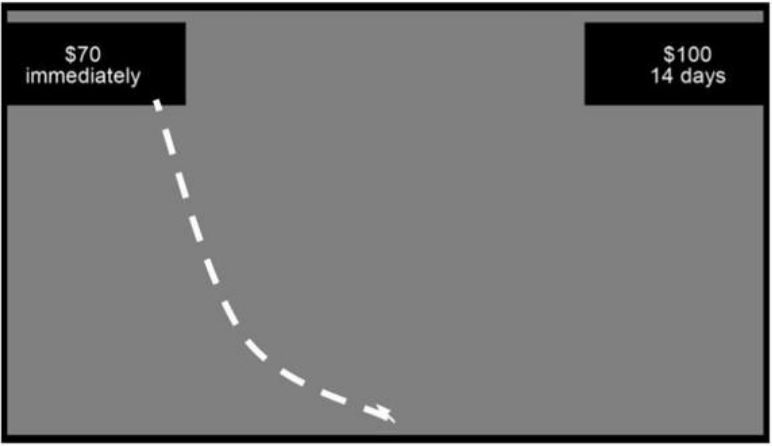

*Note:* the response is made by moving a mouse and clicking on a response box.

Participants completed 90 trials in total, comprised of 80 trials representing all possible combinations of rewards and delays, and ten control trials in which both options were presented at no delay. Participants were required to make a selection by clicking on a button with options drawn on the screen. The positioning of an option on either left or right was randomly chosen at the beginning of each trial.

The discounting rate was calculated for each participant in accordance with the hyperbolic model:

$$V = \frac{A}{(1 + kD)^{\beta}}$$

In this model, *A* is the amount of reward, *V* is the subjective value of a reward, *D* is the delay associated with the reward, *β* is the choice temperature parameter (indicating choice consistency), and *k* is an individual's discounting rate (intertemporal impatience; Stillman & Ferguson, 2019). The Bayesian delay discounting model was chosen because it allows for a more precise estimation of the discounting rate than traditional maximum likelihood estimation methods (Vincent, 2016). The R package *hBayesDM* (Ahn et al., 2017) was used to estimate the discounting rate.

**Mouse movement measures**. Mouse cursor movement measures were shown to be indicative of impulsive action and impulsive choice tendencies (Leontyev & Yamauchi, 2019, 2021; Leontyev et al., 2018). In the stop-signal task, in addition to traditional RT/accuracy-based measures, we selected maximum velocity, maximum acceleration, total distance, and stopping distance of mouse movement as initial candidates for behavioral indicators. The peak velocity of movement is known to provide insights into how we value things, specifically the subjective value of choice we make (Shadmehr & Ahmed, 2020). Mouse-cursor movement features, such as the area under the curve (AUC), has also been associated with the uncertainty of the decision-maker (Stillman et al., 2018; Stillman & Ferguson, 2019). Evidence suggests that AUC and maximum velocity time measured in a delay discounting task corresponds to concrete components of decision making, such as decision thresholds and non-decision times in a stop-signal task (Leontyev & Yamauchi, 2021; Ratcliff & McKoon, 2008).

Total distance was calculated by summing up the shortest distances between each of the successively recorded coordinates. Maximum velocity was estimated as the largest result of dividing each distance between successively recorded coordinates $d$ ($d_1$, $d_2$, $d_3$ … $d_n$) by recording time $t$ (~16 ms). Maximum acceleration was calculated as the largest difference between velocities on two adjacent segments $d$ on the trajectory (e.g., $\frac{d_2}{t} - \frac{d_1}{t}$). The area under the curve (AUC) was calculated by subtracting the area below the ideal shortest distance between the starting point and a response button from the area above the shortest distance. Finally, the stopping distance was equal to the total distance that a cursor traveled after a stop-signal was delivered. Calculation of mouse movement measures is illustrated in Figure 3. Due to the significant skewness and anomalies of the mouse-cursor movement features, the raw scores from these measures were converted using a rank-based inverse normal transformation as suggested by Bishara and Hittner (2012).

As a measure of inattentiveness, we employed standard deviation (SD) of maximum velocity in "go" trials in the stop-signal task as a mouse movement equivalent to the response time (Leontyev & Yamauchi, 2021). Performance in "go" trials (in the stop-signal and related tasks) is representative of one's attentional control (Bocharov et al., 2021). For example, omission errors in the Continuous Performance task are one of the best indicators of attentional deficits in ADHD (Berger et al., 2017). Likewise, "go" errors in a go/No-go task are indicative of inattentiveness among children (Bezdijan et al., 2009). Response times in a random dot kinematogram reflect an individual's visual attention capacity (Hanning et al., 2019).

In the delay discounting task, in addition to the discounting rate and choice consistency parameter, measures collected in DDT included maximum velocity, maximum acceleration, total distance, and area under the curve (AUC) of mouse movement. These measures were computed separately in trials where participants selected sooner-smaller and larger-later options (Figure 3).

**Figure 4**

*Calculation of the mouse movement measures*

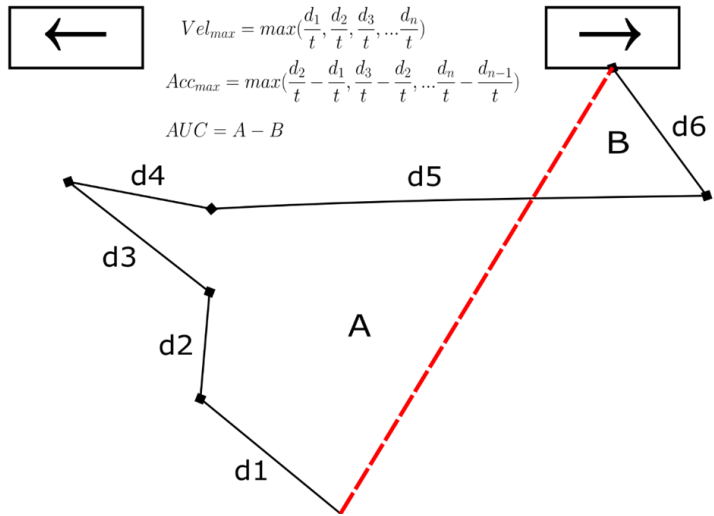

*Note:* The red dashed line shows the shortest path between the starting point and the response button. d1, d2, d3… d6 denote the shortest distances between successively recorded coordinates. AUC = area under the curve. All measures were calculated using an R package *mousetrap* (Kieslich & Henninger, 2017).

After completing the stop signal and delay discounting task, all participants were presented with the Urgency-Premeditation-Perseverance-Sensation Seeking-Positive Urgency scale (UPPS), Conners Adult ADHD Scale (CAARS), and the Barratt Impulsiveness Scale (BIS-11). Additionally, we included the Emotional Reactivity Scale (ERS) and the Schizotypal Personality Questionnaire (SPQ) at the end to gather data for separate, unrelated pilot studies

**Urgency-Premeditation-Perseverance-Sensation Seeking-Positive Urgency scale (UPPS-P).** Urgency-Premeditation-Perseverance-Sensation Seeking-Positive Urgency scale (UPPS-P; Whiteside and Lynam, 2001) is a 59-item questionnaire designed to assess emotional impulsivity. UPPS-P has five scales: Negative Urgency, Positive Urgency (assess tendencies to act rashly under negative/positive emotions), Lack of Premeditation (assesses the tendency to act without forethought), Lack of Perseverance (assesses the ability to remain focused on a task), and Sensation Seeking (assesses the tendency to seek out novel experiences). Consistency measures (Cronbach's α) range from .83 to .89 for the four subscales. The questionnaire presents statements to which participants respond by indicating their agreement or disagreement on a scale from 1 to 4, with 1 being "agree strongly" and 4 being "disagree strongly," e.g., "I have a reserved and cautious attitude toward life." In our sample, the reliability estimates ranged from .83 to .81

**Conners' Adult ADHD Rating Scales (CAARS).** Impulsivity is a core component of the attention deficit/hyperactivity disorder (Winstanley et al., 2006). Comorbid ADHD might exacerbate impulsive choice and impulsive action tendencies; to control for this possibility, we used Conners' Adult ADHD Rating Scales (CAARS). Conners Adult ADHD questionnaire – self-report long version (CAARS-S: L) is a widely accepted ADHD assessment tool. CAARS is a 66-item measure that asks participants to indicate how accurately the questionnaire's statements describe participants' personal feelings from the past two weeks until the present time. Responses are coded on a scale of 0 to 3. Higher scores represent a statement's stronger indication of a participants' current condition. Internal consistency estimates for CAARS range from .79 to .90 for all subscales (Conners et al., 1999). In our sample, reliability estimates were in the range .95 – .98.

**Barratt Impulsiveness Scale (BIS-11).** As a self-report measure of impulsive choice and impulsive action, we used the Barratt Impulsiveness Scale (BIS-11; Patton et al., 1995). It has 30 items, organized into three subscales: Attentional (BIS-A), Motor (BIS-M), and Nonplanning (BIS-N) impulsivity. The items are descriptions of an individual, e.g., "I make-up my mind quickly." A participant must indicate how often a given statement describes them on a scale of 1 to 4, with 1 being "Rarely/Never" and 4 being "Almost Always/Always." Internal consistency estimates typically range from .79 to .83 (Reid et al., 2014); in our sample, Cronbach's $\alpha$ was equal to .89.

**Results**

The initial step in assessing "Core Impulsive Action" and "Core Attentional Impulsivity" theories of emotional impulsivity involves identifying the behavioral indicators that are associated with the four latent constructs of impulsivity—impulsive action, impulsive choice, attentional impulsivity, and emotional impulsivity. To achieve this, a confirmatory factor analysis (CFA; Jacobucci et al., 2019) was performed to identify behavioral measures that cluster under the four latent factors.

**Latent factors in the "Core Impulsive Action" model.** First, we tested the "Core Impulsive Action" model, which postulates that emotions primarily impact an individual's ability to suppress motor reactions. A three-factor model incorporating Impulsive Action (IA), Impulsive Choice (IC), and Emotional Impulsivity (EI) was fitted to the mouse movement data.

In addition to the questionnaire measures of impulsive action (namely, Barratt Impulsiveness Scale Motor subscale, BIS-M), several behavioral measures of response inhibition were initially used as indicators of impulsive action: commission errors (Broos et al., 2012) and stop-signal reaction time (Verbruggen & Logan, 2008). In addition, stopping distance – a mouse movement measure of impulsive action well-correlated with other impulsivity measures (Leontyev et al., 2019) – was also included as an indicator. The initial model that included all four (SSRT, BIS-M, stopping distance, and commission error) measures showed a poor fit: $\chi^2(24)$ = 65.81, $p$ <.05; SRMR = .07; CFI = .90; RMSEA = .09. After removing SSRT as an indicator, the model fit improved. The final model incorporated BIS-M scores, commission error, and stopping distance: $\chi^2(17) = 33.40$, $p = .01$; SRMR = .05; CFI = .96; RMSEA = .07).  In this final

model, stopping distance showed a weak correlation ($r = 0.09$) with the impulsive action latent

construct, while commission error exhibited a stronger correlation with impulsive action (IA)

latent construct ($r = 0.33$).

Initial indicators of impulsive choice included two commonly used indices of impulsive behavior, discounting rate and choice consistency parameter β, as well as BIS-Nonplanning subscale scores. Following Dschemuchadse and colleagues, the area under the curve was also used as a mouse-cursor movement measure of impulsive choice (Dschemudchadse et al., 2013).

This model, however, showed a poor fit: $\chi^2(24) = 124.12$, $p < .05$; SRMR = .11; CFI = .81;

RMSEA = .14. After removing BIS-Nonplanning scores, the fit improved: $\chi^2(17) = 33.40$, $p$

= .01; SRMR = .05; CFI = .96; RMSEA = .07. Here area under the curve (AUC) of the mouse movement in the delay discounting task was highly correlated with the latent construct of impulsive choice (IC).

Finally, Emotional Impulsivity factor included UPPS-Negative Urgency and UPPS-

Positive Urgency scores as indicators. The final three-factor (IC, IA, EI) model indicated an acceptable fit: $\chi^2(17) = 33.40$, $p = .01$; SRMR = .05; CFI = .96; RMSEA = .07, BIC = 4184.28,

with average factor loadings ranging from .09 to .83 (mean = .20). The final "Core Impulsive

Action" measurement model is illustrated in Figure 3.

**Figure 5**

*Core Impulsive Action model*

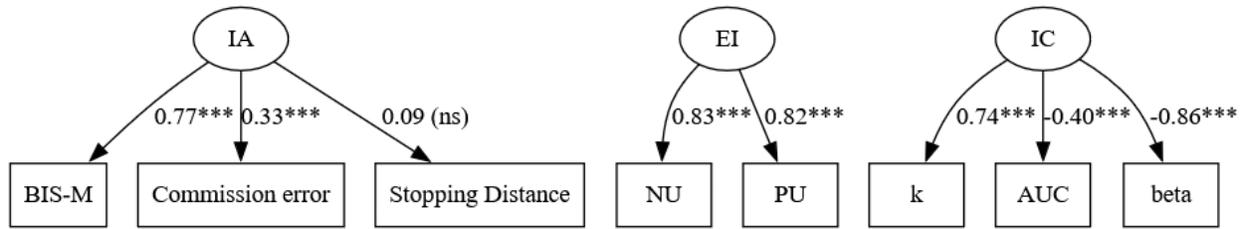

*Note*: IC = Impulsive Choice, EI = Emotional Impulsivity, and IA = Impulsive action. k =

discounting rate in the delay discounting task, beta = choice consistency parameter (β) in DDT

(delay discounting task), AUC = area under the curve of the mouse movement in the delay

discounting task. NU/PU = Negative/Positive Urgency in UPPS. BIS-M = BIS-Motor

Impulsivity, Commission error = 1 - proportion of making a response in "stop" trials in the stop

signal task, and Stopping Distance = distance travelled after the stop-signal was delivered in

SST.

**Latent factors in the "Core Attentional Impulsivity" model**. The "Core Attentional

Impulsivity" model stipulates that emotions primarily impact an individual's ability to control and direct attention. This model was largely based on the "Core Impulsive Action" model described above, with the addition of the Attentional Impulsivity (AImp) latent construct. EI, IC,

and IA latent constructs remained the same as obtained in the final "Core Impulsive Action"

model.

We initially selected BIS-Attentional Impulsivity scores, CAARS-Inattention scores as

the major indices of attentional impulsivity and included additional behavioral measures taken

from the stop signal task - random dot kinematogram direction discrimination accuracy

(Felisberti and Zanker, 2005), mean maximum velocity and mean maximum acceleration in "go"

trials (a measure of inattention in the random dot kinematogram; Levy et al., 2018), and standard deviations of maximum velocity and acceleration in "go" trials (Hauser et al., 2016).

The model with direction discrimination accuracy and acceleration in "go" trials in the stop signal task showed a poor overall fit: $\chi^2(48) = 122.26$, $p < .05$; SRMR = .08; CFI = .91; RMSEA = .09. After excluding direction discrimination accuracy and replacing acceleration with velocity, the final four-factor (IC, IA, EI, AImp) model showed a good fit: $\chi^2(38) = 67.09$, $p < .05$; SRMR = .06; CFI = .96; RMSEA = .06, BIC = 5594.23. The average factor loadings ranged from 0.09 to 0.87 (mean = 0.15). The latent construct of attentional impulsivity was dominated by the two questionnaire measures, BIS attentional subscale and CAARS attention subscale ($r\,\dot{s} = 0.83$). Behavioral measure, maximum velocity in "go" trials in the Stop Signal Task was weakly correlated ($r = 0.13$) with the latent construct of attentional impulsivity. The final "Core Attentional Impulsivity" model is illustrated in Figure 6.

**Figure 6**

*Core Attentional Impulsivity model*

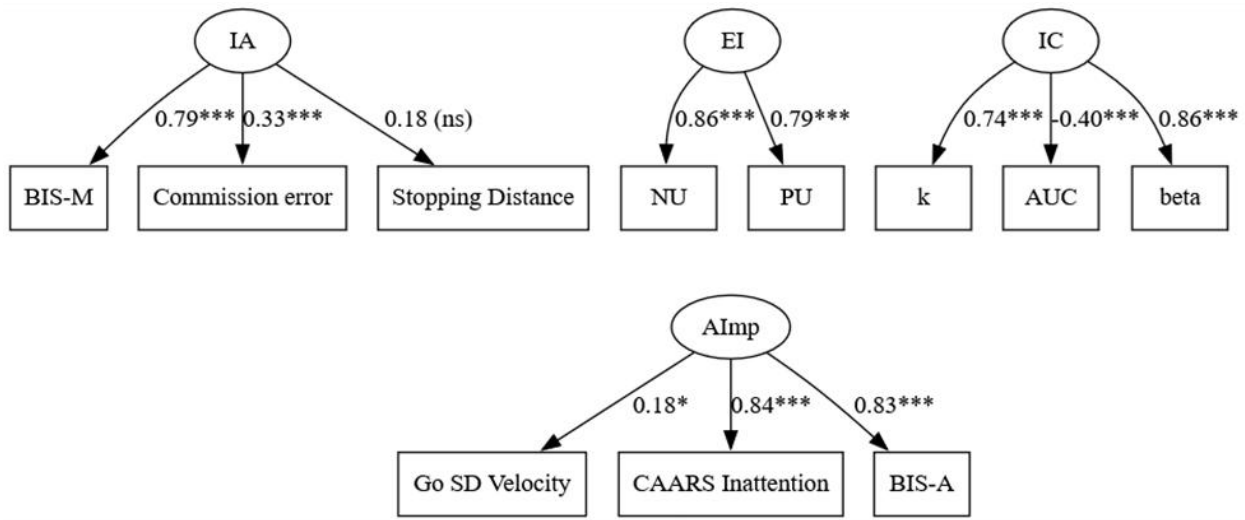

*Note*: IC = Impulsive Choice, EI = Emotional Impulsivity, IA = Impulsive action, AImp = Attentional Impulsivity. k = discounting rate, AUC = area under the curve of mouse motion, and beta = choice consistency parameter (β) in the delay discounting task (DDT). NU/PU = Negative/Positive Urgency in UPPS. BIS-M = BIS-Motor Impulsivity, Commission error = proportion of making a response in "stop" trials in the stop signal task (SST), and Stopping Distance = distance travelled after the stop-signal was delivered in SST. Go SD Velocity = standard deviation of maximum velocity in "go" trials in the stop signal task (SST), CAARS Inattention = CAARS Inattention subscale, and BIS-A = BIS-Attentional Impulsivity scale.

**Structural Equation Modeling: Core Impulsive Action versus Core Attentional Impulsivity models**

Is impulsive action or attentional impulsivity a better predictor of emotional impulsivity and impulsive choice? To answer this question, we compared structural equation models that correspond to "Core Impulsive Action" and "Core Attentional Impulsivity" accounts of impulsivity. As in confirmatory factor analysis, we employed R package *lavaan* (Rosseel, 2012).

***"Core Impulsive Action" model***. The "Core Impulsive Action" model is equivalent to the model proposed by Madole et al. (2020). This model places impulsive action (poor response inhibition) as the predictor for emotional impulsivity, which in turn predicts impulsive choice. If this model is sound, impulsive action should predict emotional impulsivity, and emotional impulsivity should predict impulsive choice (Figure 7).

The fit indices of the "Core Impulsive Action" model indicated an acceptable fit: χ2(18) = 33.49, *p* =.02; SRMR = .05; CFI = .96; RMSEA = .07, illustrated in Figure 7. Contrary to the predictions, the link between emotional impulsivity and impulsive action was not significant (*β* = 0.95 *p* = .19), as was the link between impulsive action and impulsive choice (*β* = 0.45, *p*=.19). Altogether, these findings suggest that the "Core Impulsive Action" model fell short in accounting for the covariance structure of impulsive action, emotional impulsivity, and impulsive choice measures.

**Figure 7**

*Final "Core Impulsive Action" model.*

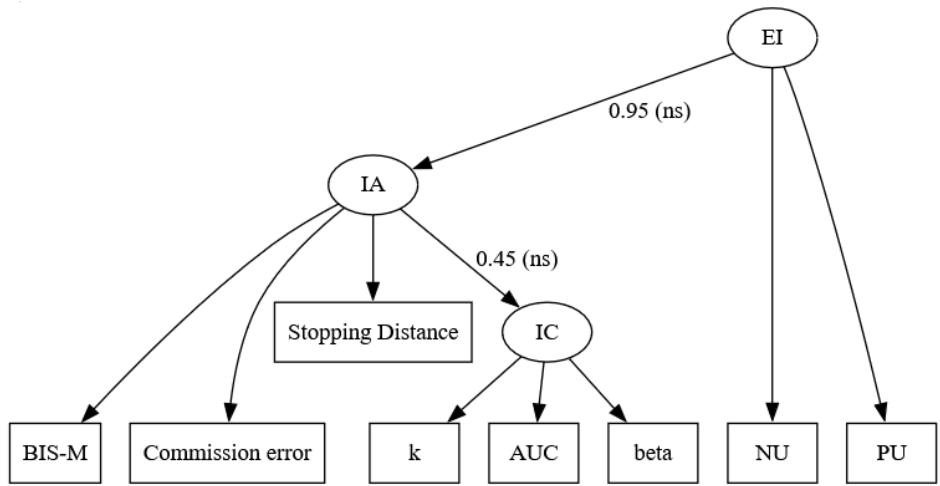

*Note*: ns = not significant.

**<u>"Core Attentional Impulsivity" model</u>**. The "Core Attentional Impulsivity" model places emotional impulsivity as the predictor of attentional impulsivity and attentional impulsivity as the predictor of impulsive action and impulsive choice. This model is different from the "Core Impulsive Action" model in that attentional impulsivity plays a central role for the general construct of impulsivity (emotional impulsivity, impulsive choice, and impulsive action). If this model is sound, emotional impulsivity should predict attentional impulsivity, and attentional impulsivity should predict both impulsive choice and impulsive action.

Consistent with the prediction, emotional impulsivity significantly predicted attentional impulsivity, which in turn predicted impulsive choice tendencies. Attentional impulsivity did not significantly predict impulsive action; instead, contrary to the initial hypothesis, impulsive action was correlated with emotional impulsivity.

Fit indices of the final "Core Attentional Impulsivity" model showed an acceptable fit: $\chi^2(40) = 70.87$, $p < .05$; SRMR = .06; CFI = .96; RMSEA = .06. The final (revised) model is illustrated in Figure 8. In sum, structural equation modeling confirmed the predictions of the "Core Attentional Impulsivity" model. Altogether, our data favors the "Core Attentional Impulsivity " model.

**Figure 8**

*Final "Core Attentional Impulsivity" model.*

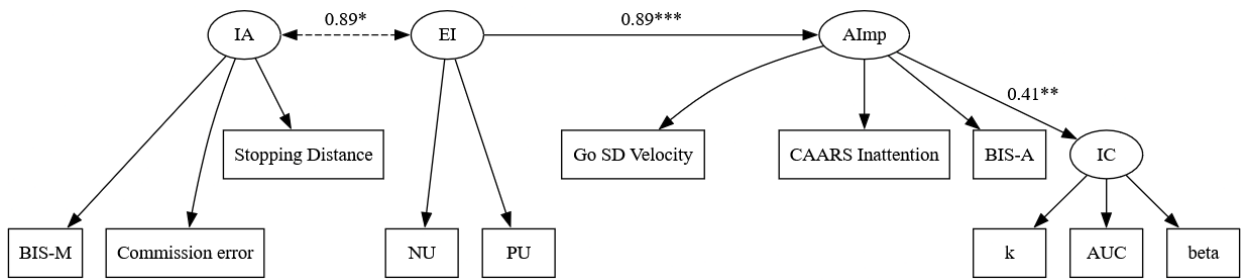

Note: *p <.05, ** p<.01, ***p <.001. Dashed lines indicate paths that were changed or added to the original "Core Attentional Impulsivity" model.

**Interim conclusions**

Consistent with past literature (Scherbaum et al., 2016), impulsive choice is best represented by delay discounting rate ($k$), consistency parameter ($\beta$), and area under the curve of the mouse cursor movement identified in the delay discounting task. Impulsive action is represented by an individual's accuracy in inhibiting the response when a stop signal is present (commission errors) and the average distance an individual's cursor traveled after the stop signal was delivered (stopping distance). Finally, attentional impulsivity was represented by the standard deviation of maximum velocity in "go" trials. In Study 2, we used these behavioral indices to assess the "Core Impulsive Action" and "Core Attentional Impulsivity" theories.

Structural equation modeling showed the advantage of the Attentional Impulsivity model ("Core Attentional Impulsivity") over the Response Inhibition ("Core Impulsive Action") model. As the results show, the predictor of emotional impulsivity is the inability to sustain attention, evidenced by significant relationships between attentional impulsivity, emotional impulsivity, and impulsive choice. Parallel to Madole et al. (2020), these results also suggest that deficits in response inhibition (impulsive action) are not a predictor of emotional impulsivity but rather are a correlate of emotional impulsivity (Figure 7).

**Study 2: Emotion Elicitation and Resulting Changes in Mouse-cursor Motions**

The purpose of Study 2 was to experimentally test the "Core Impulsive Action" and "Core Attentional Impulsivity" models. We elicit positive, negative, and neutral emotions in participants (within-subjects design) and investigate whether elicited emotions modify behaviors (mouse-motion patterns) pertinent to impulsive action, impulsive choice, or attentional impulsivity. As indicators of attentional impulsivity, impulsive action, and impulsive choice, we used the behavioral measures confirmed by Study 1 to represent the aforementioned constructs (Table 1).

Study 2 consisted of two sessions: neutral and emotional, one week apart. In the emotional session, participants viewed pleasant or unpleasant pictures during the stop-signal and delay discounting tasks (Figure 9). In the neutral session, participants viewed neutral pictures during the stop-signal and delay discounting tasks. We used nearly the same SST and DDT, as in Study 1, with minor changes: the participants had emotions elicited using standardized emotional pictures (the International Affective Picture System, IAPS, and Open Affective Standardized Image Set, OASIS; Bradley & Lang, 2017; Kurdi et al., 2017)

**Table 1**

*Constructs and corresponding measures*

| Construct | Questionnaire measures | Behavioral measures |
|---|---|---|
| Impulsive action | BIS-Motor | Commission error, stopping distance. |
| Impulsive choice | - | Discounting rate, area under curve, consistency parameter |
| Attentional Impulsivity | CAARS-Inattention (subscale A), BIS-Attention | SD of maximum velocity in "go" trials |
| Emotional Impulsivity | UPPS-Negative Urgency (NU), UPPS-Positive Urgency (PU) | - |

**Figure 9**

*Example of task sequence of SST and DDT task blocks and PANAS-X in the emotional condition.*

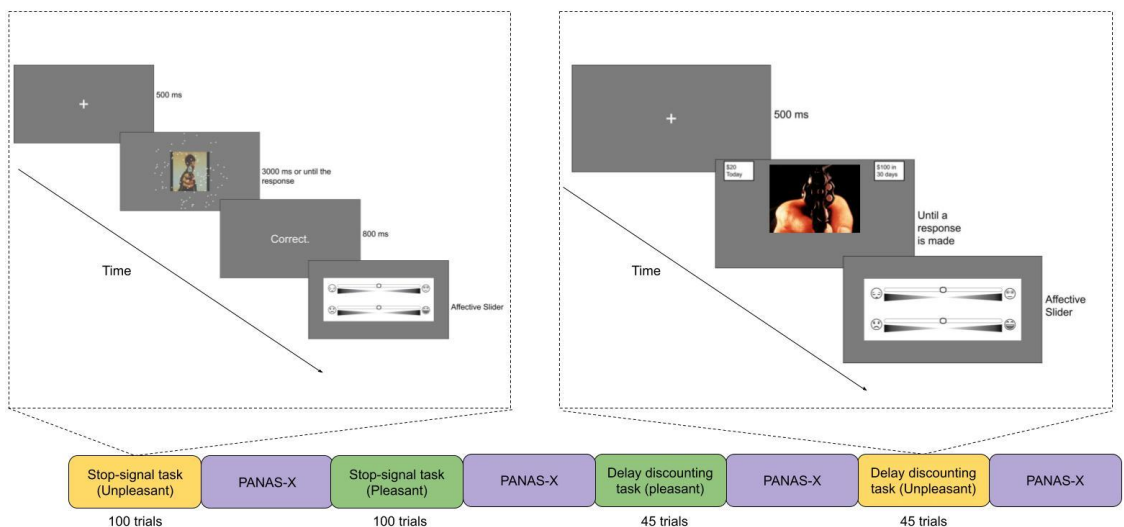

*Note:* in this example, the participant first completed SST block with unpleasant pictures, followed by SST block with pleasant pictures. In the delay discounting task, the block with pleasant pictures was delivered first, followed by the block with unpleasant pictures. All participants completed SST first, then DDT; the order of pleasant/unpleasant blocks was randomly determined for each participant. In the neutral condition, all emotional stimuli blocks contained neutral pictures.

**Predictions**

We employed contrast analysis (Rosenthal & Rosnow, 1985) and investigated the extent to which positive, negative, and neutral emotions influence impulsive action and attentional impulsivity.

If the core deficit in emotional impulsivity is impulsive action—reduced ability to inhibit undesirable action due to heightened emotions, participants' SST performance (impulsive action) should deteriorate in trials that embed emotional stimuli compared to emotionally neutral stimuli. That is, participants should show longer stopping distance and more commission errors in the emotional condition. Alternatively, if the core deficit in emotional impulsivity is inattention— reduced ability to maintain attention due to heightened emotions, we expect significantly higher inattention measures (i.e., more variability in maximum velocity in go trials) in the emotional condition.

If positive and negative emotions selectively influence impulsive action and attentional impulsivity, high or low impulsive action and attention impulsivity metrics should emerge depending on the valence of emotional pictures. In particular, we expect that unpleasant pictures result in longer stopping distance and pleasant pictures produce high variability in maximum velocity in go trials (Figure 10).

**Figure 10**

*Predictions of "Core Impulsive Action" and "Core Attentional Impulsivity" of emotional*

*impulsivity*

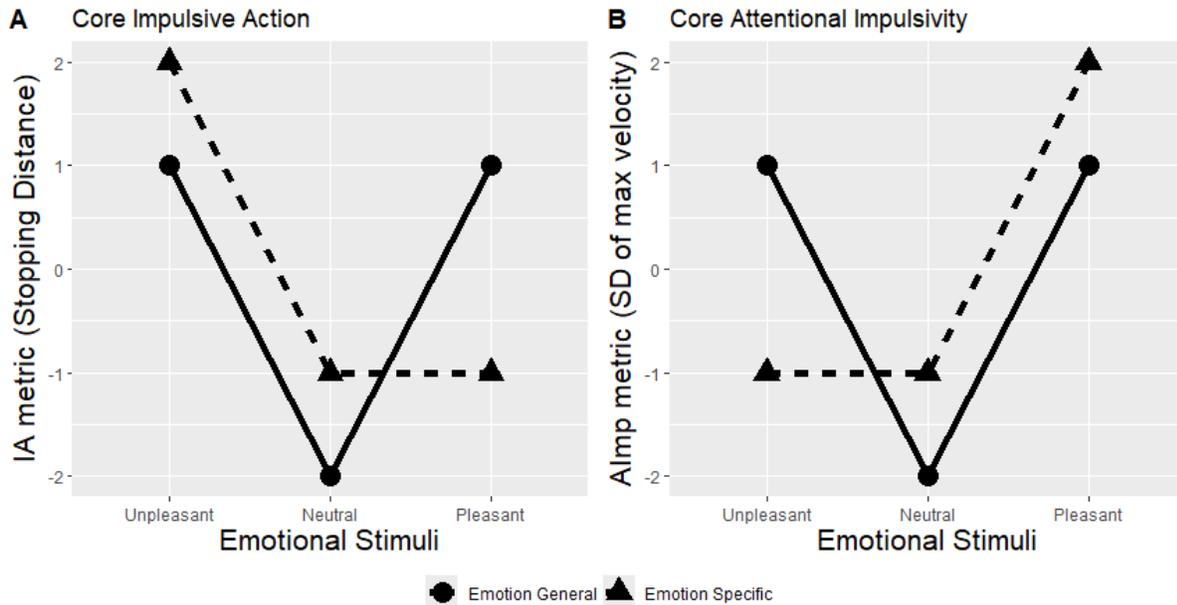

*Note*: Predictions by Core Impulsive Action account (emotions disrupt response inhibition – the ability to inhibit undesirable action) – (a), and Core Attentional Impulsivity account (emotions disrupt the ability to maintain attention) – (b). If the impact of emotion is general, both unpleasant and pleasant emotional pictures should elevate Impulsive Action (IA) and Attentional Impulsivity (AImp) metrices (solid lines). If the impact is emotion-specific, unpleasant and pleasant emotional pictures should selectively modify Impulsive Action (IA) and Attentional Impulsivity (AImp) metrices (dashed lines). The numbers on the y-axes represent contrast analysis weights assigned to the respective metrics.

## Methods

### Transparency and Openness

In the sections that follow, we report how we determined the appropriate sample size, exclusions, all manipulations and all measures in the study. All data are available at https://osf.io/wgu6p/?view_only=72958acaba264690bbf2a2100349672d/. Data were analyzed using R, version 4.2.3 (R Core Team, 2022). Study 2's design and analyses were not pre-registered. This study has been approved by the Institutional Review Board of Texas A&M University (approval ID: IRB2021-0908D).

### Participants

Based on *a priori* power analysis conducted using Gpower (Faul et al., 2007), it was determined that a minimum of 74 individuals would be required to achieve an effect size f = 0.5 with a power of 0.95 for the repeated measures ANOVA with three measurements. Consequently, we recruited a total of 78 individuals from a university subject pool in exchange for course credit. Forty-eight of these individuals indicated their gender as female and 30 as male. We excluded the participants who failed to achieve  a maximum of 95% of commission errors, and 5% direction discrimination accuracy in the stop signal task either in the emotional or neutral condition. We chose this threshold because including more trials improves the reliability of inhibitory indices (e.g., SSRT; Congdon et al., 2012), particularly when healthy adult samples are considered[2]. The final sample consisted of 61 individuals (mean age 18.79, SD = 1.17), with 35 females (mean age 18.8, SD = 1.37) and 26 males (mean age 18.8, SD = 0.85).

### Procedure

The experiment had two sessions (emotional and neutral), one week apart. All participants took part in both sessions, both spanning about one hour. Each session had two blocks of trials with different background pictures: pleasant/unpleasant (emotional session), or neutral (neutral session, illustrated in Figure 9). In the emotional condition, pleasant or unpleasant pictures were shown in the background of each trial (Figure 9) and in the neutral condition, emotionally neutral pictures were shown in the background. The order of sessions and blocks was randomized for each participant.

---

[2] Similar to Exp. 1, we also applied more stringent criteria (20 and 40% primary task and stop-signal task accuracy). The results of these analyses remained largely the same.

During the emotional session, participants engaged in two tasks: the Stop-Signal Task (SST) and the Delay Discounting Task (DDT). Each task block featured either pleasant or unpleasant images in the background. Following the completion of each block, participants responded to the PANAS-X questionnaire (Watson and Clark, 1994). Additionally, they completed the UPPS-P questionnaire at the conclusion of the session. The neutral session mirrored the emotional session, with the only difference being the presentation of neutral images in the background. An example of the task sequence can be found in Figure 9.

In the emotional session, participants first carried out the stop-signal task (200 trials total). The task had two blocks: one block (100 trials) contained pleasant pictures and the other (100 trials) contained unpleasant pictures. Each block was followed by the Positive and Negative Affect Schedule – Expanded (PANAS-X) questionnaire (Watson and Clark, 1994**).** Following the stop signal task, participants completed the delay discounting task (90 trials, presented two blocks of 45 trials), and each block was followed by the PANAS-X questionnaire. After completing SST, DDT, and PANAS-X, participants completed the UPPS-P questionnaire (Figure 9). The neutral condition was identical to the emotional condition except that all emotional pictures were replaced with neutral pictures. The scores on Positive and Negative Urgency were in the ranges 1.33 – 3.50 and 1.23 – 3.38, respectively.

**Stop-signal task.** In the stop-signal task, individual trials were presented in two blocks of 100 trials. Among those, 80% of the trials were "go" trials and 20% were "stop" trials. The stop signal task was identical to the one described in Study 1 except that emotional or neutral pictures were shown in the background. As in Study 1, participants were presented with a random dot kinematogram (100 dots) and judged the left-right direction of moving dots. At the end of each trial, participants rated their arousal and valence using Affective Slider (Figure 9). At the end of each 100-trial block, participants were asked to complete the Positive and Negative Affect Schedule questionnaire (PANAS-X; illustrated in Figure 9).

**Delay discounting task.** We modified Study 1's delay discounting task to include emotional and neutral blocks of 45 trials each. In emotional blocks, pleasant or unpleasant pictures were presented in the background; in neutral blocks, neutral pictures were presented in the background. After each trial, participants rated their arousal and valence with Affective Slider; after each block, they completed PANAS-X (see Figure 9).

*Materials*

**Emotional stimuli.** The study employed a combination of emotionally evocative and neutral images obtained from two sources: the International Affective Picture System (IAPS) and the Open Affective Standardized Image Set (OASIS, Kurdi et al., 2017). The IAPS had 1195 pictures, while the OASIS had 900, as the IAPS did not provide enough highly stimulating stimuli for each stop-signal task trial.

The images in both the IAPS and OASIS sets are rated based on two dimensions: valence and arousal. The ratings for these dimensions were normalized with a mean of 0 and standard deviation of 1. For the emotional condition, images with arousal ratings above 0.9 and valence ratings above 1 or below -1 for pleasant and unpleasant trials respectively were chosen. For the neutral condition, images with valence ratings below 1 and above -1 were selected. Overall, 580 images were chosen, with 290 for the emotional condition and 290 for the neutral condition.

To evaluate the effects of emotional stimuli on arousal and valence, two self-report measures (PANAS-X and Affective Slider) were employed.

**Positive and Negative Affect Schedule – Extended Form (PANAS-X).** Positive and Negative Affect Schedule – Extended Form (Watson and Clark, 1994) is a 60-item self-report questionnaire designed to measure positive and negative affect. The scales consist of different phrases and words (e.g., "dissatisfied with self") that describe feelings and emotions that a participant might have experienced. Participants rate how accurately a word or phrase describes their feeling from 1 (very slightly or not at all) to 5 (extremely). PANAS-X measures two higher-order scales (Positive and Negative Affect) and 11 specific affects: Fear, Sadness, Guilt, Hostility, Shyness, Fatigue, Surprise, Joviality, Self-Assurance, Attentiveness, and Serenity. Internal consistency coefficients range from .83 to .90 for the Positive Affect scale and from .84 to .91 for the Negative Affect scale.

**Affective slider (AS).** Affective Slider (Betella and Verschure, 2016) is a computer-based self-report tool that allows one to quickly assess an individual's subjective pleasure and arousal associated with an emotional stimulus via two digital sliders. Affective slider has been validated to produce results similar to the Self-Assessment Manikin (SAM; Betella and Verschure, 2016). AS' reliability estimates range from 0.87 to 0.93 (Imbault et al., 2018).

### *Design*

This study utilized a within-subjects design to test the "Core Impulsive Action" and "Core Attentional Impulsivity." Contrast analysis was conducted to tease apart the predictions made by the two theories (Figure 10).

Moderation analyses were performed to further test the two theories. If the"Core Impulsive Action" theory is correct, emotional impulsivity should moderate the relationship between impulsive action and impulsive choice. Conversely, if the "Core Attentional

Impulsivity" theory is valid, emotional impulsivity should moderate the relationship between attentional impulsivity and impulsive choice.

**Results**

*Manipulation Check*

We first examined whether the emotion elicitation method was effective. This analysis, illustrated in Figure 11, showed that 1) emotional stimuli produced immediate change in affect; and 2) this change was not limited to a given trial and persisted throughout SST and DDT.

**Figure 11**

*Manipulation Check*

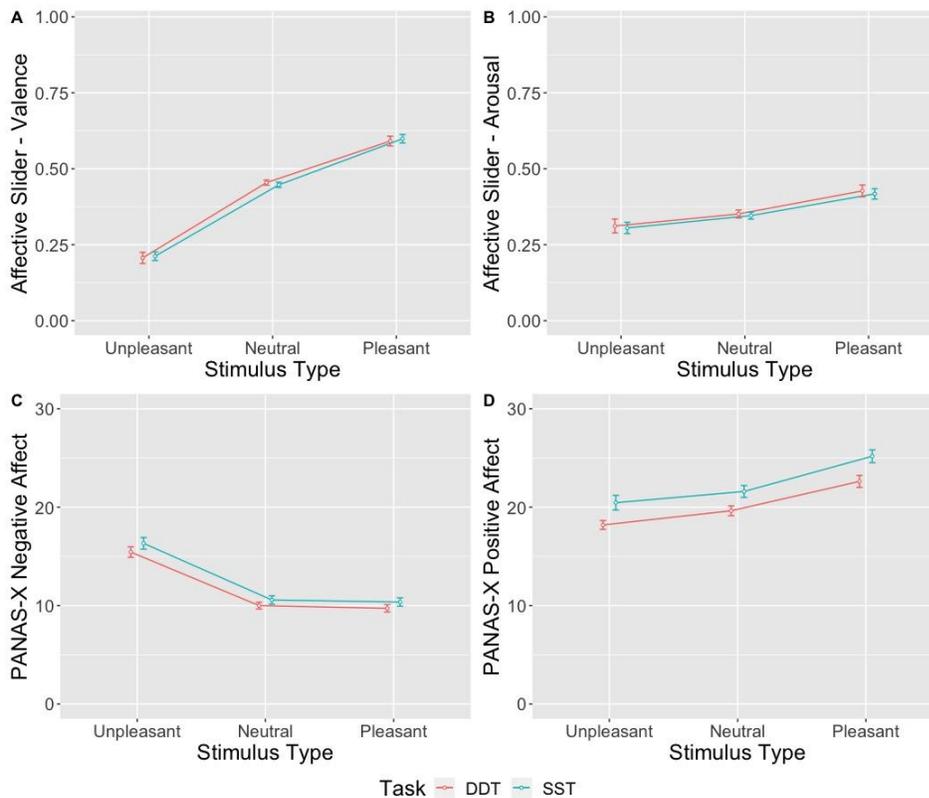

*Note*: A 3x2 repeated measures ANOVA indicated that emotional elicitation was effective: (A) Emotion slider–- Valence shows that people reported higher valence (i.e., more pleasant feeling) immediately after viewing pleasant pictures than unpleasant or neutral; (B) Likewise, participants indicated that they felt more aroused after pleasant pictures compared to neutral or unpleasant. Furthermore, the influence persisted through SST and DDT: (C) Participants reported higher negative affect after the SST/DDT blocks with unpleasant pictures, compared to blocks with neutral or pleasant pictures; (D) Participants reported higher positive affect after the SST/DDT blocks with pleasant pictures, compared to blocks with neutral or unpleasant pictures. Points represent the average scores for a given stimulus type (pleasant, unpleasant, and neutral) and task (SST/DDT); error bars represent the within-subject standard error of the mean.

**Valence.** To ensure that participants' affect changed as a result of looking at pictures, we conducted a 3x2 (Task: SST/DDT vs. Pleasant/Neutral/Unpleasant) repeated-measures ANOVA with valence ratings (PANAS and Affective Slider) as dependent measures. The results show that, as expected, participants reported high positive affect after observing pleasant pictures and high negative affect after observing unpleasant pictures. Moreover, the impact of the pictures lasted throughout SST and DDT. Given Affective Slider, we found a significant main effect of stimulus type (pleasant/neutral/unpleasant) ($F$ (2,120) = 193.25, $MSE$ = 0.02, $p$ <.001, $\eta^2$ = .57), but no significant effect of task (SST vs DDT; $F$ < 1.0) nor interaction between task and stimulus type ($F$ < 1.0).

Given PANAS-X, blocks containing pleasant stimuli resulted in higher positive affect, while blocks with unpleasant stimuli led to higher negative affect: positive affect, $F$ (2,120) = 18.11, $MSE$ = 37.46, $p$ <.001, $\eta^2$ = .05; negative affect, $F$ (2, 120) = 53.23, $MSE$ = 25.04, $p$ <.001, $\eta^2$ = .234. We also found a significant main effect of task in both blocks: blocks containing pleasant stimuli, SST vs DDT, $F$ (1,60) = 25.23, $MSE$ =18.5, $p$ <.001, $\eta^2$ = .02; blocks containing unpleasant stimuli, SST vs DDT, $F$ (1, 60) = 7.27, $MSE$ = 6.16, $p$ =.009, $\eta^2$ =.004). There was no interaction between task and stimulus type ($F$ < 1.0).

**Arousal.** On the Affective Slider – Arousal scale, participants rated pleasant pictures as more arousing than neutral or unpleasant pictures, regardless of the task ($F (2, 120) = 5.32$, $p =.007$, $\eta^2 = 0.05$). Post-hoc comparisons with Tukey correction, collapsing across tasks (SST/DDT), revealed that trials with pleasant pictures elicited significantly higher arousal than those with neutral ($t(60) = 3.58$, $p = .002$) or unpleasant pictures ($t(60) = 3.64$, $p = .002$), while there was no significant difference in arousal between neutral and unpleasant stimuli ($t (60) = 1.67$, $p = .22$). No significant main effect of task ($F < 1.0$) or interaction between task and stimulus type was observed ($F < 1.0$).

In summary, our results showed that the emotional pictures influenced participants' valence and arousal ratings. However, it is important to note that unpleasant stimuli were rated as less arousing. The emotional effect persisted throughout both tasks, as evidenced by higher PANAS-X Positive Affect scores after SST/DDT blocks with pleasant pictures, and higher PANAS-X Negative Affect scores after SST/DDT blocks with unpleasant pictures. Additionally, Positive Affect was higher after SST than after DDT, which may be due to participant exhaustion later in the experiment.

### *Main analyses*

Parallel to Study 1, before testing "Core Impulsive Action" and "Core Attentional Impulsivity" theories, we transformed behavioral measures (stopping distance (IA), commission error (IA) and SD of maximum velocity in "go" trials (AImp)) using rank-based inverse normal transformation (Bishara & Hittner, 2012). The transformation was applied simultaneously to the metrics collected in emotional and neutral conditions.

**Figure 12**

*Results of the contrast analysis*

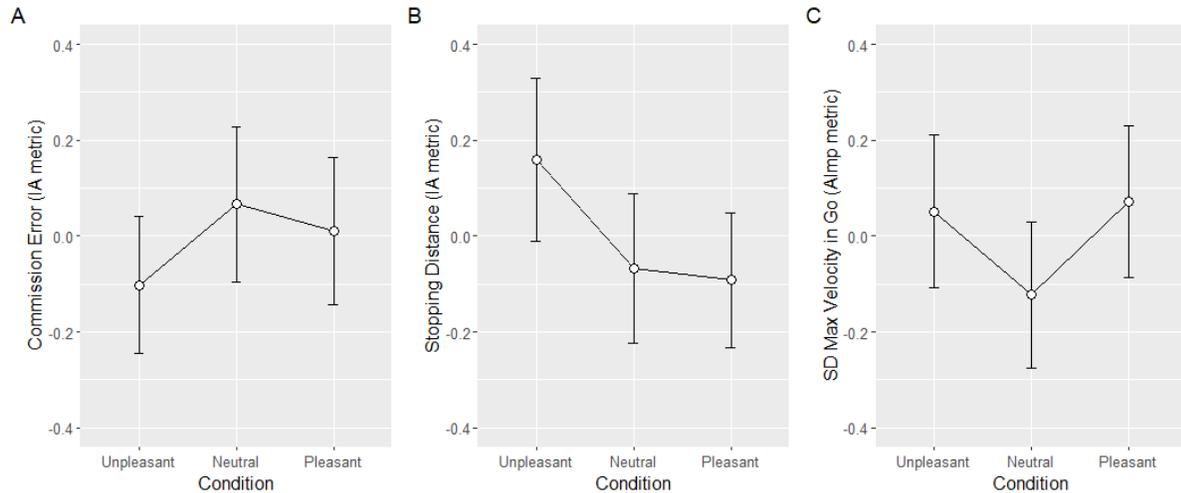

*Note*: Results of contrast analysis for Core Impulsive Action account (A and B), and Core Attentional Impulsivity account (C). Impulsive action metrics were higher in unpleasant condition; Attentional Impulsivity metrics were higher in unpleasant and pleasant, compared to neutral, conditions. Error bars represent within-subject standard error of the mean.

**Did emotional stimuli affect impulsive action (disrupt response inhibition) or attentional impulsivity (disrupt attention)?** First, we tested the "Core Impulsive Action" theory. Specifically, we tested whether emotional (pleasant and unpleasant) pictures produced more impulsive actions (reduce response inhibition) than neutral pictures. We entered commission error and stopping distances as dependent variables. Our contrast analysis suggests that impulsive action was most likely elicited by negative emotion (unpleasant pictures) but not positive emotion (pleasant pictures). For commission error, neither a V-shaped trend ($t$ (120) = 0.93, $p$ = .36) nor a L-shaped trend ($t$ (120) = 1.17, $p$ = .24) were significant (Figure 12A). For stopping distances, we found a strong L-shaped emotion-specific trend in impulsive action metric (stopping distance) ($t$(120) = 2.58, $p$ = 0.01; Figure 12B) whereas the emotion-general V-shaped trend was absent; $t$ (120) = 1.09, $p$ = .2 .

For attentional impulsivity, we found the opposite effect. While V-shaped emotion-general trend (Figure 12C) was present for attentional impulsivity metrics (SD of maximum velocity; $t$ (120) = 2.0, $p$ = 0.048), the L-shaped emotion specific trend was lacking in the attentional impulsivity metric ($t$(120) = 1.17, $p$ = 0.24). These findings suggest that both positive and negative emotions produced attentional impulsivity.

**Moderation analysis.** According to "Core Impulsive Action" and "Core Attentional Impulsivity" theories, impulsive action and attentional impulsivity lead to impulsive choice when emotions are stirred, and emotional impulsivity (positive / negative urgency as measured by UPPS) is said to exacerbate impulsive choice. Here, we employed moderation analysis and examined the extent to which emotions deteriorate impulsive choice.

If "Core Impulsive Action" view is valid, then the behavioral indicator of impulsive action, stopping distance (Table 1), should correlate with the behavioral indicator of impulsive choice, delay discounting rate. If the "Core Attentional Impulsivity" theory is valid, the behavioral indicator of attentional impulsivity, the standard deviation of maximum velocity in "go" trials, should correlate with impulsive choice. Moreover, this relationship should be pronounced for participants with high emotional impulsivity as measured by UPPS in the emotion condition (pleasant or unpleasant pictures are shown in the background).

To test this hypothesis, we performed a multiple parallel moderation analysis using PROCESS macro (Hayes, 2022) and investigated the extent to which emotion impulsivity (high UPPS scores), impulsive action (high/low stopping distance), attentional impulsivity (high/low SD of maximum velocity in "go"), and impulsive choice (delay discount rates) interact.

*"Core Impulsive Action" model.* Emotional impulsivity (NU and PU scores, obtained from UPPS questionnaire) did not moderate the relationship between impulsive action and impulsive choice. Stopping distance, NU, and PU scores did not account for a significant amount of variance in the impulsive choice measure (delay discount rate $k$): $F_{(5,55)} = 1.22$, $p = .31$, $MSE = 0.98$, $R^2 = .10$. No significant interactions between stopping distance and either PU or NU were detected (PU x stopping distance: $\beta = 0.11$, $p = .51$, illustrated in Figure 13A); NU x stopping distance: $\beta = -0.001$, $p = .99$, (Figure 13C). We conducted the same moderation analysis separately for the positive and negative emotion conditions. This additional analysis did not result in any significant moderation effect; pleasant stimuli condition, $F_{(5, 55)} = 1.18$, $p = .33$, $MSE = 0.98$, $R^2 = .10$; unpleasant emotion condition, $F_{(5, 55)} = 1.41$, $p = .23$, $MSE = 0.97$, $R^2 = .11$.

Commission error showed poor performance as well. Neither commission errors, nor NU or PU scores accounted for a significant amount of variance in the delay discounting rates in emotional ($F_{(5,55)} = 1.29$, $p = .28$, $MSE = 0.97$, $R^2 = .10$) or neutral ($F_{(5,55)} = 0.61$, $p = .68$, $MSE = 1.03$, $R^2 = .05$) conditions.

*"Core Attentional Impulsivity" model.* Emotional impulsivity (positive urgency PU and negative urgency NU) significantly moderated the relationship between attentional impulsivity and impulsive choice measures. SD of maximum velocity in "go" trials, NU, and PU scores together accounted for a significant amount of variance in delay discounting rate $k$, $F_{(5, 55)} = 2.90$, $p = .02$, $MSE = 0.86$, $R^2 = .21$. An interaction term between PU and SD of maximum velocity in "go" trials scores significantly added to the amount of accounted variance in delay discounting rates: $\Delta R^2 = .06$, $\Delta F_{(1, 55)} = 4.48$, $\beta = 0.33$, $t_{(55)} = 2.12$, $p = .04$ (Figure 13B). Interaction between NU and SD of maximum velocity in "go" trials did not significantly add to the amount of explained variance in discount rates: $\Delta R^2 = .005$, $\Delta F_{(1, 55)} = 0.39$, $\beta = -0.08$, $t_{(55)} = -0.63$, $p = .53$ (Figure 12D). This relationship is summarized in Table 2. The moderation effect was present in the pleasant but not unpleasant emotion conditions: pleasant emotion, $F(5, 55) = 3.81$, $p = .005$, $MSE = 0.81$, $R^2 = .26$; unpleasant emotion, $F(5, 55) = 1.36$, $p = .25$, $MSE = 0.97$, $R^2 = .11$

**Table 2**

*Conditional effects of the SD of maximum velocity in "go" trials at values of NU and PU*

| NU value | PU value | $\beta$ | 95% CI | SE | $t$ | $p$ |
|---|---|---|---|---|---|---|
| Mean – 1SD | Mean – 1 SD | 0.07 | [-0.34,0.47] | 0.20 | 0.34 | 0.73 |
| Mean – 1SD | Mean | 0.36 | [-0.07, 0.78] | 0.21 | 1.69 | 0.10 |
| **Mean – 1SD** | **Mean + 1SD** | **0.76** | **[0.09, 1.41]** | **0.33** | **2.27** | **0.03** |
| Mean | Mean – 1SD | -0.04 | [-0.38, 0.31] | 0.17 | -0.20 | 0.84 |
| **Mean** | **Mean** | **0.25** | **[0.01, 0.50]** | **0.12** | **2.09** | **0.04** |
| **Mean** | **Mean + 1SD** | **0.64** | **[0.18, 1.10]** | **0.23** | **2.80** | **0.01** |
| Mean + 1SD | Mean – 1SD | -0.12 | [-0.63, 0.38] | 0.25 | -0.49 | 0.63 |
| Mean + 1SD | Mean | 0.16 | [-0.20, 0.53] | 0.18 | 0.91 | 0.36 |
| **Mean + 1SD** | **Mean + 1SD** | **0.55** | **[0.11, 0.99]** | **0.22** | **2.52** | **0.01** |

*Note*: significant effects are in the boldface type

**Figure 13**

*Probing the relationship among Stopping distance/ SD of maximum velocity in "Go" trials,*

*Positive/Negative Urgency (PU/NU), and discounting rate (k).*

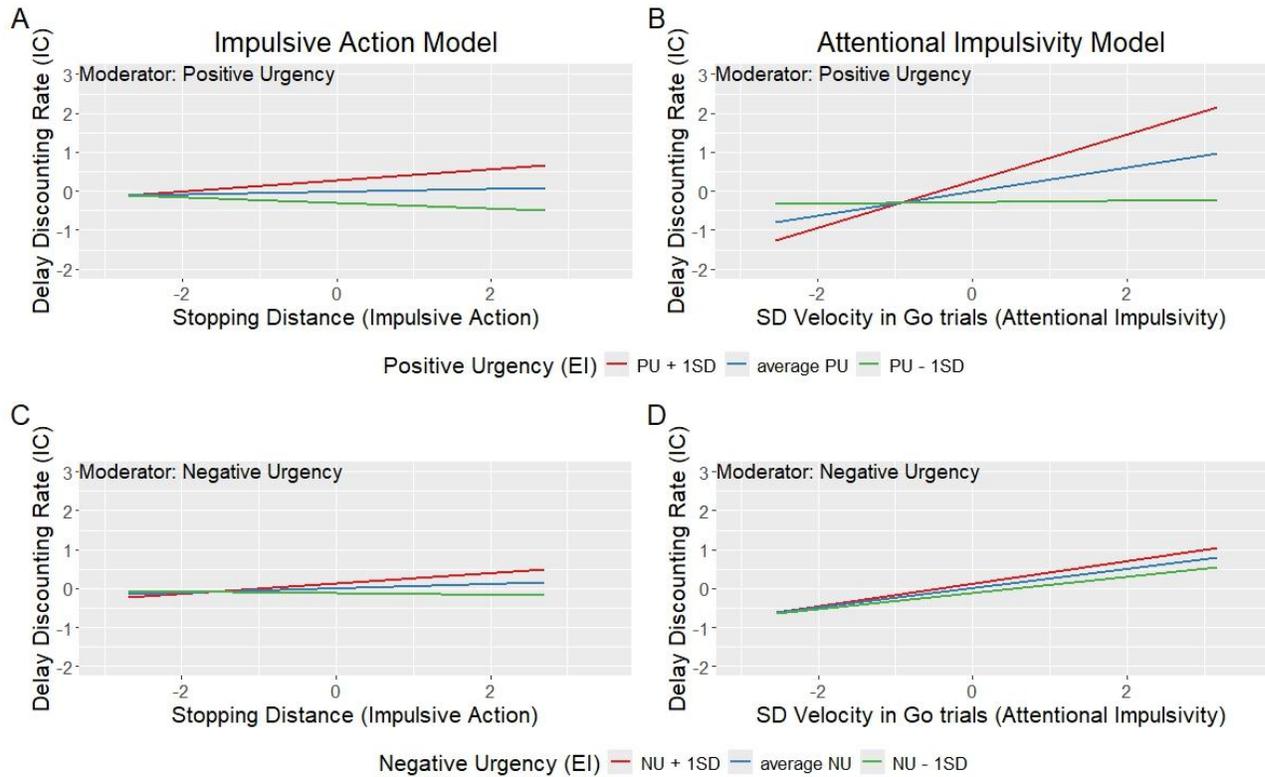

*Note*: panels A and B show positive urgency as moderator; panels C and D show negative urgency as moderator.

In sum, the relationship between impulsive choice (delay discounting rate *k)*, emotional

impulsivity (NU and PU in UPPS), and attentional impulsivity (SD of maximum velocity in "go"

trials) can be described as follows. Levels of attentional impulsivity (as measured by SD of maximum velocity) predicts impulsivity choice (as measured by delay discounting rates k) of participants with moderate (mean) and high (mean + 1SD) levels of Positive Urgency (illustrated in Figure 13). Negative Urgency does not play a role in the relationship between SD of maximum velocity in "go" trials and delay discounting rates.

## Discussion

### Summary of findings

In this study, we investigated whether the emotions primarily increase impulsive behavior by boosting impulsive action or by disrupting attentional control. Our contrast and moderation analyses suggest that, in a nonclinical population, attentional impulsivity is linked to emotional states: both positive and negative emotions induce attentional impulsivity. Moreover, we found that attentional impulsivity interacts with emotional impulsivity, specifically, positive urgency. Such an interaction was lacking in the case of impulsive action.

The results of our study are consistent with the "Core Attentional Impulsivity" account, which posits attentional control as a main route of emotions' influence. Standard deviation of maximum velocity in go trials were higher in the emotional condition, suggesting higher inattention in the presence of emotional stimuli. Much like the "Core Attentional Impulsivity" theory predicted, Positive Urgency also moderated the relationship between impulsive choice and inattention in the emotional condition. That is, individuals who displayed poor attentional control also showed high impulsive choice, but this interaction was largely limited to those who displayed high tendencies to act impulsively under the influence of positive emotions.

Traditionally, the research on impulsivity focused on negative emotions as drivers of maladaptive impulsive behaviors, such as gambling (Navas et al., 2017), food addiction (Parylak et al., 2011) or drug abuse (Dorison et al., 2020). For example, Parylak and colleagues show that food-seeking behavior is often triggered by the negative stress. Similarly, Navas and colleagues show that impulsive behaviors in gambling disorder is associated with stronger negative emotions. Finally, Dorison and colleagues displayed sadness as the most important predictor of different substance use behaviors (e.g., number of puffs taken by a smoker as well as impatience for a cigarette).

However, recent work has underscored that positive emotions can also stimulate impulsive behaviors. As a case in point, Vintro-Alcaraz and colleagues (Vintro-Alcaraz et al., 2022) show that the severity of gambling disorders is correlated with both Positive and Negative Urgency. Regarding impulsive action, although previous research linked response inhibition to specifically Negative Urgency (Allen et al., 2019), given our findings it is possible that Positive Urgency plays an equal if not more important role in affective control (a form of inhibitory control engaged in emotionally or motivationally salient situations; Allen, 2021). Interestingly, positive urgency has been shown to be related to other forms of inhibitory control, such as a tendency to engage in risky behavior (Cyders et al., 2007).

It seems that inattention is more correlated with positive but not negative emotional urgency. This lack of relationship requires further investigation; perhaps, the difference between attentional impulsivity and response inhibition (which is known to be correlated with negative urgency more than positive; Johnson et al., 2020) lies in the specific facet of emotional impulsivity they are related to. That is, specific emotions might directly affect specific types of impulsive behaviors: negative emotions impacting primarily response inhibition and positive emotions impacting the ability to focus and maintain attention. Future studies should investigate the specific effect of negative versus positive emotional stimuli, as well as their interaction with trait negative/positive emotional urgency.

Taken together, our findings highlight the importance of positive emotions in producing impulsive behavior. Traditionally, the research on impulsivity focused on negative emotions as drivers of maladaptive impulsive behaviors, e.g., gambling (Navas et al., 2017), food addiction (Parylak et al., 2011) or drug abuse (Dorison et al., 2020). Recent studies highlight that positive emotions can drive impulsive behaviors as well. Our research provides an important insight into the mechanism of the relationship between positive emotions and impulsivity, which is integral for developing optimal therapeutic and pharmaceutic interventions.

One of the significant implications of our research pertains to the understanding of impulsivity in general. Many studies examining the relationship among different forms of impulsivity (e.g., impulsive choice and impulsive action) reveal weak or even nonexistent correlations within community samples (Broos et al., 2012; MacKillop et al., 2016). This lack of agreement has led some to question the validity of a unified concept of impulsivity (see MacKillop et al., 2016 for example). Our findings, grounded in community samples, suggest that it may be premature to dismiss impulsivity as a unifying construct. Rather, it appears that impulsivity should be best understood in the context of emotions. Clinical disorders are often characterized by emotion dysregulation, which can influence an individual's ability to make sound decisions, inhibit motor responses, and maintain attentional control. However, in community samples, the "normal" everyday level of emotional influence, unamplified by the clinical disorder, might simply be insufficient to disturb impulse control abilities as measured by behavioral tasks. As demonstrated in Study 2, the relationship between different forms of impulsivity becomes detectable by dynamic motion-based measures of stop-signal or delay discounting tasks.

The importance of variability in attention is in line with Hauser et al. (2016): variability of responses is a more important marker of inattention than average response. However, other measures of attentional impulsivity might yield different results. One possible venue for future studies is investigating how different measures of impulsive actions (e.g., SSRT, stopping distance, antisaccades) are related to individual differences in inattention measured by other tasks, for example, the Stroop task.

Our studies highlight the potential of mouse-cursor tracking method to enhance our understanding of impulsivity. Remarkably, none of the conventional measures of impulsive behavior, even those identified by the factor analysis (e.g., stop-signal reaction time or commission error), were sensitive to emotions. In contrast, measures derived from mouse movement (such as stopping distance and the standard deviation of maximum velocity) were not only identified by the Confirmatory Factor Analysis but also exhibited at least partial responsiveness to emotions. This further validates the applicability of mouse movement tracking in impulsivity and emotion research, as confirmed in several previous studies (Leontyev et al., 2018, 2019; Leontyev et al. 2019; Yamauchi & Xiao, 2018, Yamauchi et al., 2024).

Perhaps most importantly, the effectiveness of mouse-cursor tracking provides substantial counter-evidence to the proposal by Toplak and colleagues (2013) regarding the inability of behavioral tasks to capture typical behavior (Leontyev & Yamauchi, 2021, Yamauchi et al., 2017; Yamauchi, et al., 2019). According to their view, questionnaire-based measures designed to gauge *typical*, everyday levels of impulsive behaviors (in the scope of weeks or months), contrast sharply with behavioral tasks intended to study behavior in a span of several hours, i.e., in the length of an experimental session. Moreover, behavioral tasks often prompt participants to exhibit their "best" or optimal behaviors, which have little to do with their typical behavior. That is, individuals most often do not specifically concentrate on inhibiting motor reaction for prolonged periods of time; hence, correlations between questionnaires and behavioral task measures are generally scant. Our studies, however, suggest that, with a sensitive enough measure (like mouse-cursor tracking), behavioral tasks can indeed probe trait-level impulsivity.

**Limitations**

The current study has several limitations. First, the complete scope of emotional influence may not be accurately represented within a laboratory setting. While both IAPS and OASIS are reliable and extensively tested methods for inducing emotional reactions, real-world emotional experiences are often complex and enduring, which may be challenging to reproduce in a controlled experimental setting. Moreover, our participant pool comprised undergraduate students. As college students are often pre-selected via admissions processes, they may form a population with lower emotional impulsivity and fewer impulsive behaviors compared to the general population (Hanel & Vione, 2016).

Second, in Study 1, we observed higher loading attributed to the questionnaire-based (but not behavioral) measures. In other words, questionnaire scores "dominated" the latent constructs. This problem can be traced to the generally higher reliability of questionnaire-based measures, as compared to the behavioral measures (Dang et al., 2020).

Another limitation of our study lies in the distinction between attentional impulsivity, impulsive choice, and impulsive action. While the present study treated these concepts as essentially different, we acknowledge that at the fundamental level attentional and motor control strongly overlap. However, it is still valid to treat IA, IC, and AImp as different constructs. First, different neural substrates underlie impulsive choice and impulsive action (Wang et al., 2016). Moreover, impulsive choice and impulsive action are expressed differently in non-human populations, such as rats (Cho et al., 2018)

The final limitation is the design of our mouse movement studies. Multiple past studies have prompted individuals to move their mouse cursor when a trial begins. In our studies, participants received no such prompt (Wirth et al (2020: BRM), Kieslich et al (2020: BRM),

Scherbaum & Kieslich (2018: BRM), Grage et al (2019: APP), Schoemann et al (2019: BRM). We deliberately used this design to ensure the ecological validity of our study, that is, to allow them to manifest their impulsive tendencies without interference. It is possible that this choice of design could change the participants' response strategies; future studies should address this limitation.

**Future studies**

Future studies should delve deeper into the relationship between inattention and impulsive choice. More accurate methods for estimating discounting rates, such as adaptive delays (Mahalingham et al., 2018), or the introduction of actual rewards could be useful strategies. An additional prospective direction could involve integrating a task explicitly designed to probe attention—such as the Stroop task (Gronau et al., 2003)—into the current battery of tasks, which includes the Stop Signal Task (SST) and Delay Discounting Task (DDT). Future studies also should test for the potential order effects.

Another avenue of future research is a specific facet of inattention that plays a role in promoting impulsive behaviors. The ability to suppress attention to irrelevant stimuli has at least two facets: resistance to distractor interference (i.e., the ability to ignore an immediately present distracting stimulus) and resistance to proactive interference (i.e., the ability to ignore traces of memory about a stimulus). Future studies should test both abilities in their relationship with emotional impulsivity, impulsive choice, and impulsive action.

Lastly, we should note that behavioral variability demonstrates one of the essential measures of impulsivity. The importance of variability in attention aligns with Hauser et al. (2016) that variability of responses is a more important marker of inattention than average response. This link should be investigated further in future studies.

**Conclusion**

Emotional impulsivity is instrumental in a variety of impulsive behaviors, yet its mechanism remains unclear. The traditional perspective posits that impulsive action is the core deficit in emotional impulsivity. This study challenges this assumption, revealing instead that both attentional impulsivity and impulsive action correspond to different facets of emotional impulsivity.